\documentclass[aps,prb,twocolumn,showpacs]{revtex4}
\usepackage{graphicx}
\usepackage{epsf}
\usepackage{amssymb}
\usepackage{array}

\begin{document}

\title{Aging behavior of spin glasses under bond and temperature perturbations from laser illumination}

\author{
R. Arai
}
\affiliation{
Department of Applied Physics and Physico-Informatics, Keio University, 3-14-1 Hiyoshi, Kohoku-ku, Yokohama-shi, Kanagawa 223-8522, Japan
}

\author{
K. Komatsu
}
\affiliation{
Department of Applied Physics and Physico-Informatics, Keio University, 3-14-1 Hiyoshi, Kohoku-ku, Yokohama-shi, Kanagawa 223-8522, Japan
}

\author{
T. Sato
}
\affiliation{
Department of Applied Physics and Physico-Informatics, Keio University, 3-14-1 Hiyoshi, Kohoku-ku, Yokohama-shi, Kanagawa 223-8522, Japan
}

\date{\today}

\begin{abstract}
We have studied the nonequilibrium dynamics of spin glasses subjected to bond perturbation,
which was based on the direct change in the spin-spin interaction $\Delta J$, using photo illumination
in addition to temperature change $\Delta T$.  
Differences in time-dependent magnetization are observed between that under $\Delta T+\Delta J$ and $\Delta T$ perturbations
with the same $\Delta T$.
This differences shows the contribution of $\Delta J$ to spin-glass dynamics through the decrease in the overlap length. 
That is, the overlap length $L_{\Delta T+\Delta J}$ under $\Delta T+\Delta J$ perturbation is less than $L_{\Delta T}$ under
$\Delta T$ perturbation.
Furthermore, we observe the crossover between weakly and strongly perturbed regimes 
under bond cycling accompanied by temperature cycling.
These effects of bond perturbation strongly indicates the existence of both chaos and the overlap length.
\end {abstract}

\pacs{75.50.Lk,75.50.Pp,75.40.Gb}

\maketitle

\section{Introduction}

Spin glass has been studied for the past several decades, but many questions about 
the nonequilibrium dynamics of spin glass still remain. 
To clarify the nature of spin glass below the transition temperature,
the aging behavior\cite{17lundgren} in the relaxation of 
magnetization in spin glasses has been actively studied.
Especially, the aging of spin glasses subjected to perturbations $\Delta X$,
such as changes in temperature $\Delta T$ and in bond interaction $\Delta J$,
has been extensively studied because it shows characteristic behaviors peculiar to spin glasses,
such as memory and rejuvenation.\cite{24jonason,25jonsson}
These behaviors were interpreted in terms of the phenomenological scaling theory, 
called the droplet model.\cite{15fisher,16fisher,18bray}
According to this theory, the memory and rejuvenation effects are explained in terms of the concept of chaos
accompanied by the overlap length $L_{\Delta X}$.\cite{18bray,15fisher}
The correlation between two equilibrium states, before and after the perturbation, 
disappears at the length scale $L$ beyond the overlap length $L_{\Delta X}$.
However, the chaotic nature appears even at weak perturbation $\Delta X$ that satisfies $L<L_{\Delta X}$. \cite{27sheffler,30sales,26jonsson}
At strong perturbation $\Delta X$ that satisfies $L>L_{\Delta X}$,
the aging effect before the perturbation is not easily removed, but the memory effect is observed.
These contradictory aspects can be explained based on the ghost domain scenario.\cite{7yoshino,27sheffler,6yoshino,8jonsson}
Thus, the crossover between a weakly perturbed regime ($L<L_{\Delta X}$) and a strongly perturbed regime ($L<L_{\Delta X}$)\cite{6yoshino,8jonsson}
should be clarified so that we can gain an intrinsic understanding of 
the rejuvenation and memory effects based on the droplet picture.
\par
So far, the experimental studies\cite{31nordblad,5sandlund,29granberg,49bert,8jonsson}
and simulations\cite{58picco,56berthier,52berthier} of spin glasses, 
have been conducted exclusively under the temperature cycling protocol.
In such an experimental protocol based on temperature change, however, 
this change inevitably affects the thermal excitation of the droplet
and thus leads to strong separation of the time scales.\cite{7yoshino}
This makes it difficult to demonstrate the existence temperature chaos.
In fact, several papers\cite{55bouchaud,56berthier,44berthier,52berthier}
claim that the rejuvenation-memory effects observed in temperature cycling
can be attributed to the differences among the length scales caused by the change in temperature. If the {\it direct} change in the bond,$\Delta J$, can be used in this kind of experiment
without the change in time scales,
it is expected that the chaotic effect and the overlap length could be more clearly analyzed.
\par
The direct change in the spin-spin interaction $\Delta J$
can be realized through the photo excitation of carriers using a semiconductor spin glass, E.g.,
Cd$_{1-x}$Mn${_x}$Te.\cite{20galazka,3nordblad,mauger,42zhou} 
The relaxation in thermoremanent magnetization and that in isothermal remanent magnetization of
Cd$_{1-x}$Mn${_x}$Te were observed under unpolarized light illumination.\cite{28smith}
Recently, we studied the aging behavior of Cd$_{1-x}$Mn${_x}$Te under photo illumination,
and showed that the $\Delta J$ contribution can be deduced by comparing the data 
under the $\Delta T+\Delta J$ and $\Delta T$ perturbations with the same $\Delta T$.
\cite{1sato}
To date, however, there has been no systematic study of aging behavior under bond perturbation.
It is essential to obtain evidence of the existence of the chaotic effect and of
the change in the overlap length expected in the droplet model\cite{15fisher,16fisher}
through the analysis of bond perturbation data.
\par
In this study we first confirm that bond perturbation 
using photo illumination affects the spin-glass dynamics.
We estimated that $\Delta J\sim 0.14\sim 0.40$K at $\Delta T=0.26$K.  
The second purpose is to clarify the characteristics of overlap length
and to specify the crossover between weakly and strongly perturbed regimes
to demonstrate the validity of the droplet picture.
  
\section{EXPERIMENTAL DETAILS}

\begin{figure}[t]
\includegraphics[width=0.8\columnwidth]{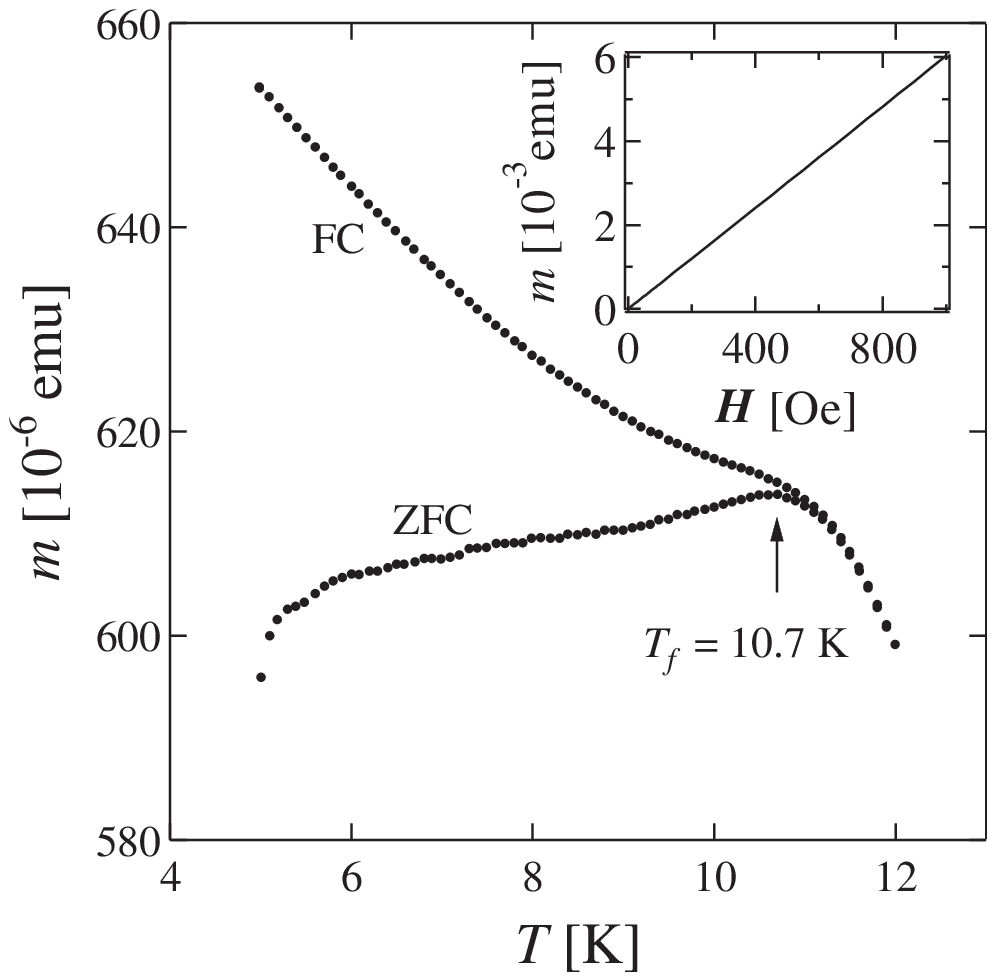}
\caption{Temperature dependence of ZFC and FC magnetizations of Cd$_{0.63}$Mn$_{0.37}$Te in
$H=100$ Oe. The freezing temperature of $\sim 10.7$K estimated by this graph.
The inset shows the field dependence of the magnetization at 7K ($=T_m$). }
\label{FCZFC} 
\end{figure}

The sample was a single crystal of Cd$_{0.63}$Mn$_{0.37}$Te (band gap energy $E_g=2.181$eV)
that was prepared using the vertical Bridgeman technique.
The magnetization of the sample, which was a plate 1.2 mm thick, was measured by a Quantum Design
MPMS5 superconducting quantum interference device (SQUID) magnetometer. Figure \ref{FCZFC}
shows the temperature dependent magnetization under FC (field-cooled) and ZFC (zero-field-cooled)
conditions in $H=100$Oe.\cite{CdMnTe_spinglass}
The spin-freezing temperature $T_f \sim 10.7$K was evaluated.
Light was guided to the sample through a quartz optical fiber so as to be parallel to
an external magnetic field. The light source was a green He-Ne laser with $\lambda =543.5$nm, 
(2.281eV), where this photon energy was slightly larger than the band gap energy of the sample. 
One side of the sample was coated with carbon. If light was illuminated on the carbon-coated side, 
the light was absorbed in a carbon black body and, thus only the thermal contribution $\Delta T$ appeared.
When the light was illuminated on the opposite side, a change in bond interaction $\Delta J$
appeared in addition to $\Delta T$.\cite{4kawai}
We determined the sample temperature during the illumination based on 
the change in field-cooled magnetization as shown in FIG. \ref{FCZFC}.
The photo-induced magnetization in Cd$_{0.63}$Mn$_{0.37}$Te was scarcely observed 
(less than 0.01 of the total magnetization change by the photo illumination),\cite{4kawai}
and thus we were able to neglect it.
The light intensity was adjusted so as to obtain the same increment of sample temperatures 
under both the  illumination conditions.
This made it possible to consider only the $\Delta J$ contribution by comparing the $\Delta T+\Delta J$ data with the $\Delta T$ data.
We note that the change in temperature by the illumination was given to a sample 
with step-like heating and cooling.\cite{4kawai} Thus we
could also neglect the effect of the finite cooling/heating rate.\cite{8jonsson,58picco}
     
\begin{figure}[t]
\includegraphics[width=0.8\columnwidth]{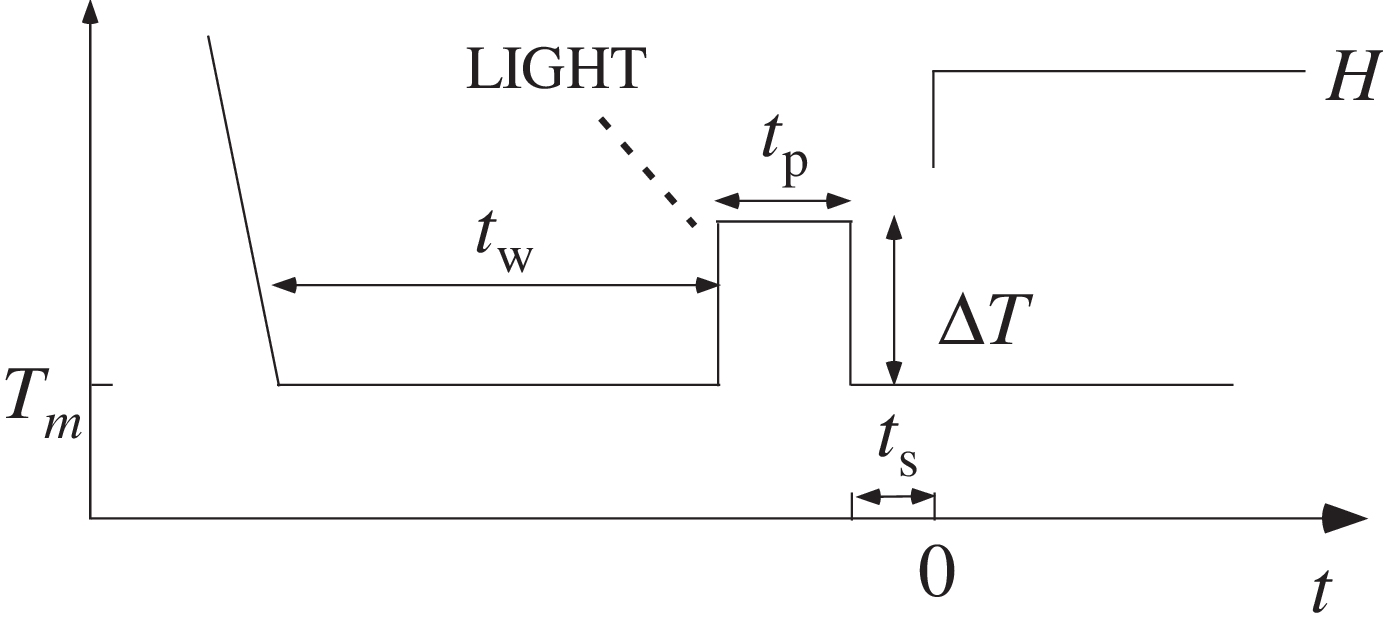}
\caption{Experimental protocol of bond-cycling process under photo illumination. 
The sample was first cooled to $T_m=7$K and aged during $t_{\rm w}=3000$s (initial aging stage). 
The perturbation was subsequently added during $t_{\rm p}$ (perturbation stage) using photo illumination.
After the perturbation was stopped, $H$ of 100 Oe was applied and 
the magnetization was measured as a function of $t$ (healing stage).}
\label{protocol} 
\end{figure}

The time dependence of magnetization was measured according to the
following procedure. (Fig. \ref{protocol}) The sample was zero-field-cooled 
down  to $T_m$ as rapidly as possible ($\sim 10$K/min)
from 30K which is above the transition temperature $T_g$. Then, the temperature $T_m$ was held
for $t_{\rm w}$ (=3000s) (initial aging stage).
After that, the perturbation was given to the sample during $t_{\rm p}$ using the photo illumination (perturbation stage). After
the light was turned off, a magnetic field $H$ of 100 Oe, at which the linear field-dependent magnetization was held
(inset of FIG. \ref{FCZFC})), was subsequently applied. 
After $t_{\rm s}\sim 60$s, the magnetic field $H$ of 100 Oe is stabilized, 
and then the magnetization $M$ was measured as a function of time $t$ (healing stage). 
\par
The dynamics of a spin glass below $T_g$ is governed by excitations of the droplet.\cite{15fisher}
The size $L_{T_m}(t)$ of the droplet, which was thermally activated at $T_m$ within the time scale $t$, 
is given by the following equation,\cite{15fisher}
\begin{equation}
L_{T_m}(t)\sim L_0{\left[\frac{k_B T_m\ln(t/\tau _0)}{\Delta (T_m)}\right]}^{1/\psi}. 
\label{droplet}
\end{equation}
Consequently, a significant difference in time scales existed even between two close temperatures,
$T_m$ and $T_m+\Delta T$.
Therefore, we define the effective duration $t_{\rm eff}(T_m)$ 
of the perturbation \cite{7yoshino} according to 
\begin{equation}
L_{T_m+\Delta T}(t_p)=L_{T_m}(t_{\rm eff}),
\label{Lteff}
\end{equation}
or
\begin{equation}
t_{\rm p}(T_m+\Delta T)=\tau _0(t_{\rm eff}(T_m)/\tau _0)^{T_m/(T_m+\Delta T)},
\label{teff}
\end{equation}
where $\tau_0$  ($\sim \hbar/J \sim 10^{-13}$s ) is a microscopic time scale.\cite{tau_value}
When we discuss the data below, we convert the actual duration of the perturbation $t_{\rm p}(T_m+\Delta T)$ into
the effective duration $t_{\rm eff}(T_m)$.

\begin{figure*}[bt]
\includegraphics[width=16cm,keepaspectratio]{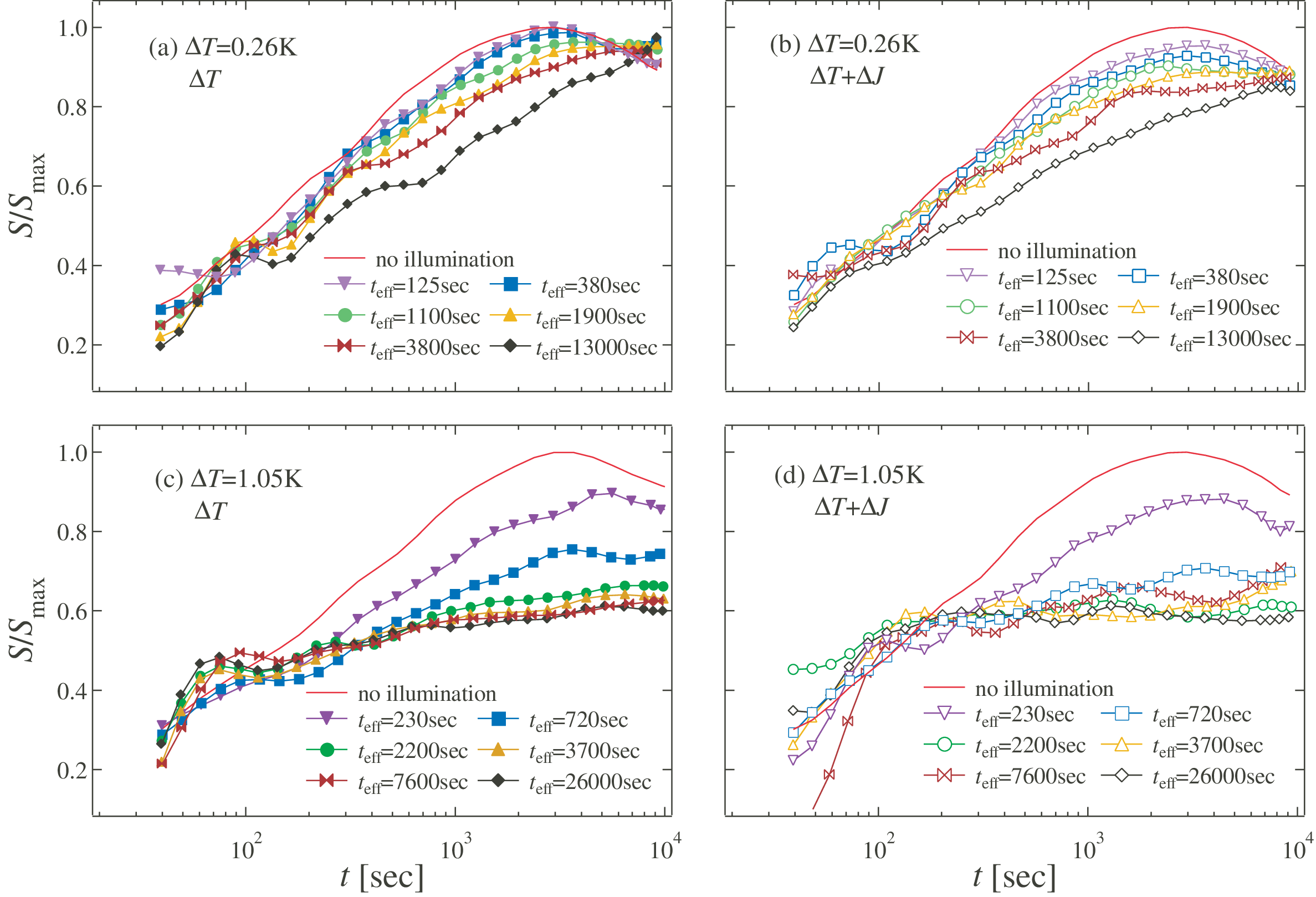}
\caption{(Color online) Time dependence of the relaxation rate $S$ at 7K. 
Solid curves show $S$, measured after the waiting time $t_{\rm w}=3000$s without illumination (isothermal aging).
Figure (a) (filled symbol) and (b) (open symbol) show the relaxation rates $S$ at various $t_{\rm eff}$ for $\Delta T=0.26$K
under $\Delta T$ and $\Delta T+\Delta J$ perturbations, respectively.
Figure (c) (filled symbol) and (d) (open symbol) show the relaxation rates $S$ at various $t_{\rm eff}$ for $\Delta T=1.05$K
under $\Delta T$ and $\Delta T+\Delta J$ perturbations, respectively.
All the data are normalized by the maximum height of $S$ without photo illumination.}
\label{S_variousteff}
\end{figure*}

\section{EXPERIMENTAL RESULTS}

We focused on the relaxation rate of the magnetization, $S=(1/H)dM/d \log t$, measured under the various conditions of strength and
duration of the perturbation. Since we observed the peaks in the perturbation time-dependent data of $S$,
the height of each peak and the corresponding peak position are important for our discussion.
\par  
The solid curves in FIG. \ref{S_variousteff} show the time dependence of the relaxation rate $S$, measured
after the aging time $t_{\rm w}=3000$s without illumination (isothermal aging).
These curves show a peak at $t \sim 3000$s, which is a typical behavior observed in many
spin glasses.\cite{18granberg}
\par
FIG. \ref{S_variousteff} also shows the typical data of perturbation time-dependent $S$ at $T_m$ (=7K) of the sample,
where the constant temperature increment $\Delta T$ (=0.26K, 1.05K) is added
during $t_{\rm eff}$ for the perturbation.
The left figures ((a) and (c)) and the right figures ((b) and (d)) show the data
under $\Delta T$ and $\Delta T+\Delta J$ perturbations, respectively. 
All the data are normalized by the maximum height of $S$ without photo illumination.
In FIGs. \ref{S_variousteff}(a) and \ref{S_variousteff}(b) ($\Delta T=0.26$K), the peak in $S$ at $t\sim t_{\rm w}$
(we call this the {\it main peak}) becomes gradually
depressed compared with the isothermal aging curve as $t_{\rm eff}$ increases.
The depression of $S$ in Fig. \ref{S_variousteff}(b)
is more pronounced than that in FIG. \ref{S_variousteff}(a) at the same $t_{\rm eff}$. 
In addition, the peak position in the main peak shifts to a longer time as $t_{\rm eff}$ increases
in both FIGs. \ref{S_variousteff}(a) and \ref{S_variousteff}(b).
In FIGs. \ref{S_variousteff}(c) and \ref{S_variousteff}(d) ($\Delta T=1.05K$),  
the main peak in $S$ is depressed with increasing $t_{\rm eff}$ 
in a more pronounced way compared with the data at $\Delta T=0.26$K.
For long $t_{\rm eff}$, $S$ becomes almost flat, but the main peak is incompletely erased.
Furthermore, in FIGs. \ref{S_variousteff}(c) and \ref{S_variousteff}(d),
a small peak appears around $t \sim 100$s (we call this the {\it sub peak}) 
except for short $t_{\rm eff}$, and
the sub-peak becomes more pronounced with increasing $t_{\rm eff}$.
The sub-peak in FIG. \ref{S_variousteff}(d) is less sensitive to $t_{\rm eff}$ 
compared with that in FIG. \ref{S_variousteff}(c).
In FIG. \ref{S_variousteff}(d), the sub-peak is so close to the main peak 
that the two peaks are insufficiently separable.
We note that some curves in $S$ in FIG. \ref{S_variousteff} show a sudden increase at large values of $t$
(e.g., S for $t_{\rm eff}\sim 7600$s in FIG. \ref{S_variousteff}(d) ($\Join$  symbols)).
This may be due to a small fluctuation in the sample temperature.

\begin{figure*}[bth]
\includegraphics[width=12cm,keepaspectratio]{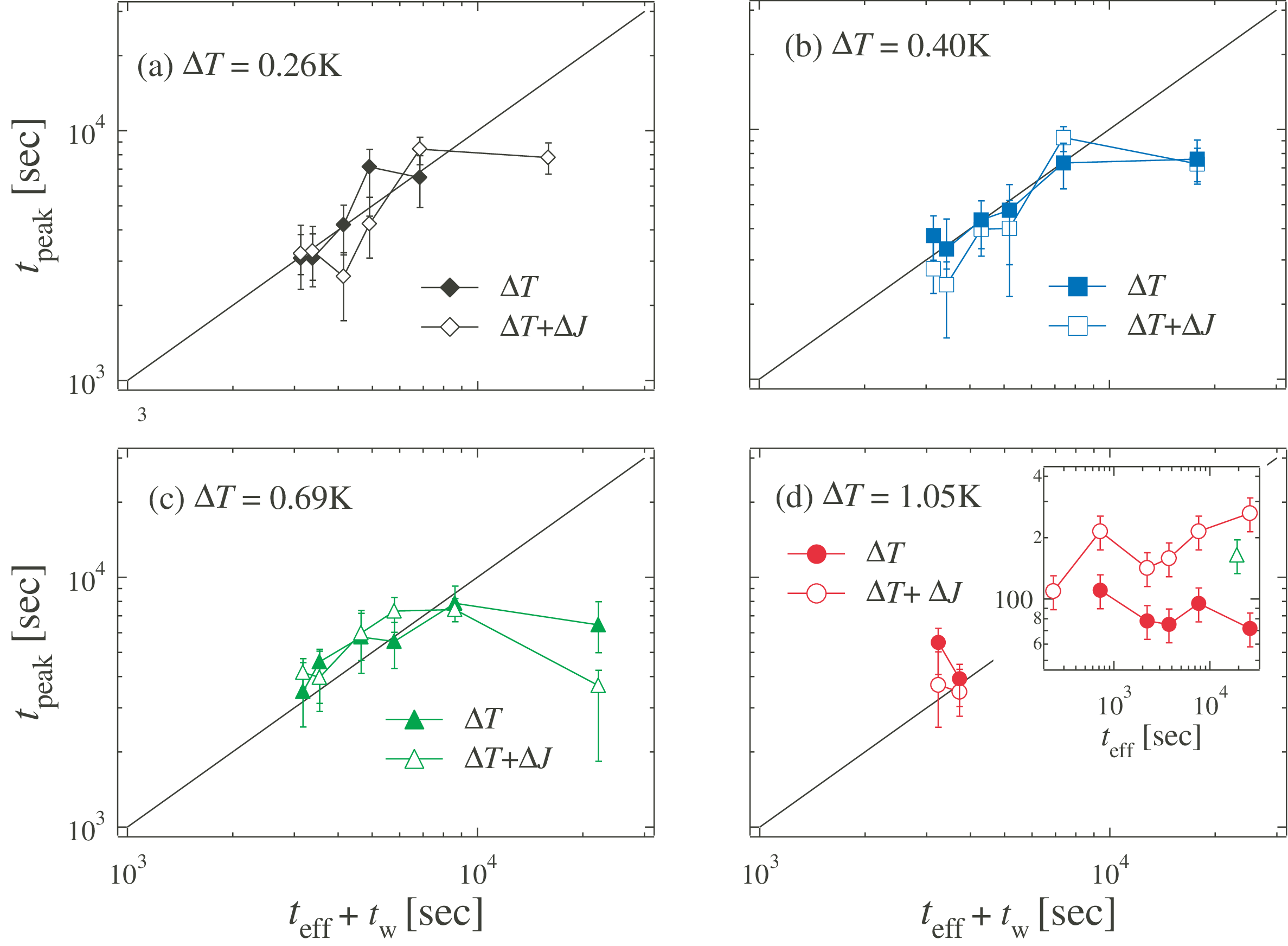}
\caption{(Color online) The $t_{\rm eff}+t_{\rm w}$ dependent $t_{\rm peak}$ at various values of $\Delta T$. 
The increments of temperature under photo illumination, $\Delta T$, are 0.26K, 0.40K, 0.69K, 1.05K.
Solid lines satisfies the equation $t_{\rm peak}=t_{\rm w}+t_{\rm eff}$.
Filled symbols and open symbols show $t_{\rm peak}$ under $\Delta T$ and $\Delta T+\Delta J$ perturbations, respectively.
The inset shows $t_{\rm eff}$-dependent $t'_{\rm peak}$. The symbols in the inset correspond to those in the main frame.}
\label{peakteff}
\end{figure*}

\section{DISCUSSION \label{discuss}}

\subsection{Theory based on the ghost domain scenario \label{theory}}
Recently, aging behavior under the bond or temperature cycle has been explained based on the behavior of domains
in terms of the ghost domain picture.\cite{7yoshino,6yoshino,8jonsson}
In the following paragraphs, we interpret our results based on the behavior of the domain under 
the bond or temperature cycle in the same way. We divide our experimental protocol into
three stages according to Fig. \ref{protocol}: the initial aging stage in which the equilibrium state $\Gamma _A$ belongs to
the environment $A-(T_m,J)$,
the perturbation stage in which the equilibrium state $\Gamma _B$ belongs to the new environment 
$B-(T_m+\Delta T,J+\Delta J)$, 
and the healing stage at the environment $A$.
\par

\begin{figure*}[t]
\includegraphics[width=15cm,keepaspectratio]{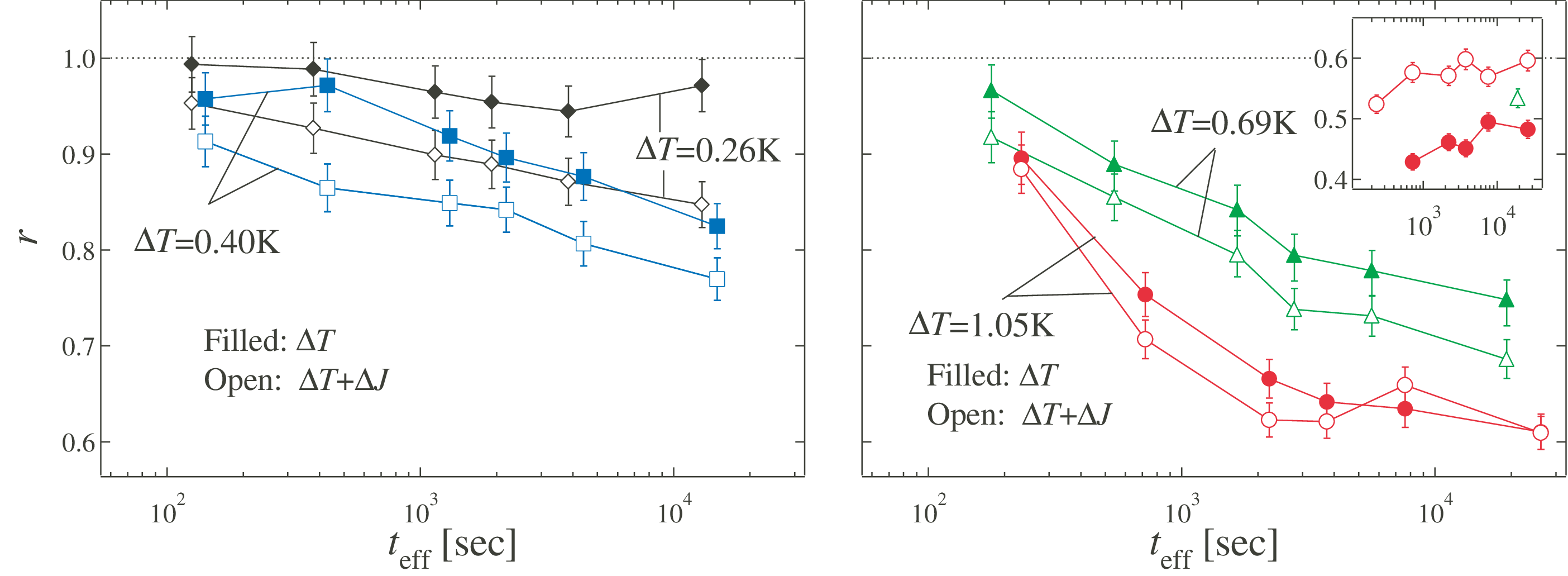}
\caption{(Color online) The $t_{\rm eff}$ dependent $r$ at various values of $\Delta T$. 
The increments of temperature under photo illumination, $\Delta T$, are 0.26K, 0.40K, 0.69K and 1.05K. 
The filled and open symbols show $r$ under $\Delta T$ and $\Delta T+\Delta J$ perturbations, respectively.
The inset shows $t_{\rm eff}$-dependent $r'$. The symbols in the inset correspond to those in the main frame.
}
\label{rteff}
\end{figure*}

In the initial aging stage, the domains belonging to $\Gamma _A$ at $T_m$ grow during $t_{\rm w}$
according to Eq. (\ref{droplet}).
In the perturbation stage, the perturbation $\Delta X$ ($\Delta T$ or $\Delta T+\Delta J$) 
is applied from $t_{\rm w}$ to $t_{\rm w}$+$t_{\rm p}$.
During the perturbation stage, the "overlap" between the equlibrium states $\Gamma _A$ and $\Gamma _B$ disappears at the length scale beyond the overlap length $L_{\Delta X}$.
The relation between $L_{\Delta X}$ and the perturbation $\Delta X$ is given as \cite{18bray,15fisher}:
\begin{equation}
L_{\Delta X}=L_0|\Delta X/J|^{-1/\zeta }.
\label{over}
\end{equation}
The chaos exponent $\zeta $ is given by $\zeta = d_s/2-\theta$ $( >0 )$,
where $d_s$ is a fractal dimension and $\theta $ is a stiffness exponent.
Theoretically, temperature and bond perturbations are equivalent 
with respect to the overlap length $L_{\Delta X}$.\cite{6yoshino}

\par
We can distinguish the weakly and strongly perturbed regimes based on 
the relationship between overlap length $L_{\Delta X}$
and the domain size grown during each stages. \cite{6yoshino,8jonsson}
If $L_{T_m+\Delta T}(t_{\rm p}) <  L_{\Delta X} $, a weakly perturbed regime appears, in which  the rejuvenation scarcely emerges.\cite{26jonsson}
If all the $L_{T_m}(t_{\rm w})$, $L_{T_m+\Delta T}(t_{\rm p})$, and $L_{T_m}(t)$ are greater than $L_{\Delta X}$, 
a strongly perturbed regime appears, in which the initial spin configuration 
is unstable with respect to droplet excitation and a new equilibrium state appears. 
This suggests a chaotic nature. 
The chaos, however, does not appear abruptly\cite{27sheffler,30sales},
and there exists a crossover between the weakly and strongly perturbed regimes.
\par
In the weakly perturbed regime, the domain belonging to $\Gamma _A$ is weakly modified  and the order parameter $\rho $ slowly decreases in the perturbation stage. 
The order parameter in the domains is easily recovered during the healing stage.
Thus, the recovery time $\tau _{\rm rec}$, which is necessary for the order parameter to saturate to 1, is given as
\begin{equation}
{\tau}_{\rm rec}^{\rm weak} \sim t_{\rm eff}.
\label{recweak}
\end{equation}
In this regime, the size of domain belonging to $\Gamma _A$ grows accumulatively so as to neglect the perturbation.
\par
In the strongly perturbed regime, the spin configuration in domains belonging to $\Gamma _A$  
is completely random
beyond the length scale of $L_{\Delta X}$, and the domains belonging to 
$\Gamma _B$ grow during the perturbation stage.
However, the effect of the initial domain of $\Gamma _A$ remains as a ghost\cite{6yoshino,8jonsson}, i.e., the domains, that grow up to $L_{T_m}(t_{\rm w})$ during initial aging, 
can vaguely keep their overall shapes (which are called {\it ghost domain}), 
and the domain interiors are significantly modified due to the growth of domains $\Gamma _B$.   
Thus, the order parameter $\rho $ significantly decreases (but does not reach zero).
When the perturbation is removed, the system recovers the initial spin configuration (healing stage). The recovery time
$\tau _{\rm rec}$ 
is given as
\begin{equation}
\tau_{\rm rec}^{\rm strong} \gg t_{\rm eff}.
\label{recstrong}
\end{equation}
In this regime, the domain belonging to $\Gamma _A$ grows non-accumulatively.

\subsection{The role of the relaxation rate under the cycling \label{theory2}}
The relaxation rate $S$ is characterized under the temperature or 
bond perturbations shown in Sec. \ref{theory}, through the peak position and height in the time dependent  $S$. 
First, we define the characteristic time $t_{\rm peak}$ corresponding to the position of the main peak.
The value of $t_{\rm peak}$ is shown as a function of $t_{\rm eff}+t_{\rm w}$ in FIG. \ref{peakteff}D
If the aging is accumulative, the spin configuration at $T_m+\Delta T$ in the perturbation stage
is equivalent to that under the isothermal aging at $T_m$ for $t_{\rm eff}$.
This results in the following relation:
\begin{equation}
t_{\rm peak}\sim t_{\rm w}+t_{\rm eff},
\label{weakpeak} 
\end{equation}
which corresponds to the reference line in FIG. \ref{peakteff}.
In the strongly perturbed regime ($L\gg L_{\Delta X}$), on the other hand, 
the chaotic nature becomes effective and the position of the main peak shifts to a shorter time compared with the accumulative aging curve.  
In this case, $t_{\rm peak}$ satisfies the relation, 
\begin{equation}
t_{\rm peak}<t_{\rm w}+t_{\rm eff}.
\label{strongpeak} 
\end{equation}
\par
During the healing stage, the order parameter of $\Gamma _A$ starts to restore the domain structure 
grown during initial aging, 
and this is probably reflected in the height of the main peak.
Thus, in order to characterize the change in the peak built in $S$, 
we define the relative peak height $r$,
which is the ratio of the height in the main peak under the perturbation to that without perturbation.
In other words, $r$ is the measure of the memory after the perturbation.\cite{8jonsson}
The value of $r$ is shown as a function of $t_{\rm eff}$ in FIG. \ref{rteff}.
\par
In the completely weakly perturbed regime ($L_{T_m}(t_{\rm eff})\ll L_{\Delta X}$), 
the order parameter fully recovers at $t\sim t_{\rm peak}$ according to Eq. (\ref{recweak}).
However, close to the strongly perturbed regime (the crossover regime between the weakly and strongly perturbed regimes),
the order parameter does not completely recover at $t=t_{\rm w}+t_{\rm eff}$.
This leads to the decrease in the height of main peak. 
In the strongly perturbed regime, the order parameter insufficiently recovers 
because of the long recovery time ${\tau}_{\rm rec}^{\rm strong}$ given in Eq. (\ref{recstrong}).
Thus, $r$ is gradually depressed as $t_{\rm eff}$ increases because of 
the rapid increase in ${\tau}_{\rm rec}^{\rm strong}$.
In addition, when the period $t_{\rm s}$ is necessary until the applied field is stabilized 
in a superconducting magnet after the perturbation is removed,
the domain grows for $t_{\rm s}$ up to the size $L_{T_m}(t_{\rm s})$.
This reflects the sub-peak.\cite{8jonsson} Thus, the sub-peak becomes pronounced as the rejuvenation effects become clear.

\begin{figure*}[tb]
\includegraphics[width=17cm,keepaspectratio]{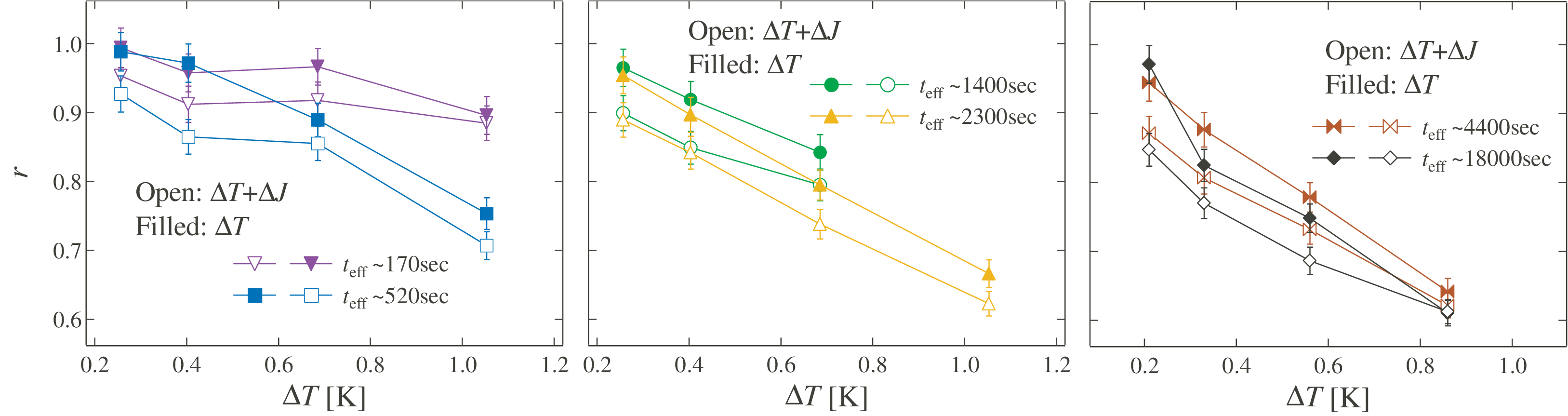}
\caption{(Color online) The $\Delta T$-dependent $r$ at various values of $t_{\rm eff}$.
Almost the same values of $t_{\rm eff}$ are gathered together in one graph.
The left, middle, and right figures show $r$ under $t_{\rm eff}\sim 170$ and $520$s, 
$t_{\rm eff}\sim 1400$ and $2300$s,
and $t_{\rm eff}\sim 4400$ and 18000s, respectively.
THe filled and open symbols show $r$ under $\Delta T$ and $\Delta T+\Delta J$ perturbations, respectively.}
\label{rDT}
\end{figure*}

\subsection{Bond-cycling experiment}
We try to classify our data, obtained under various perturbation conditions, into the following four categories: 
the weakly perturbed regime (we call this the $W$ regime);
the crossover regime, which is close to the weakly perturbed regime (we call this the $WC$ regime); 
the crossover regime, which is close to the strongly perturbed regime (we call this the $SC$ regime);
and the strongly perturbed regime (we call this the $S$ regime). 
The criteria for the classification are abridged in TABLE 1. 
The classification is mainly performed based on FIG. \ref{peakteff} and FIG. \ref{rteff} in which $t_{\rm peak}$ and $r$  are
 arranged as a function of  $t_{\rm eff}$ under various amplitudes of $\Delta T$. 
The condition, i.e., $t_{\rm eff}$ is variable and $\Delta T$ is constant, 
corresponds to the situation that the domain size $L_{T_m}(t_{\rm eff})$ 
grown during the perturbation stage is variable while the overlap lengths 
$L_{\Delta T}$ and $L_{\Delta T+\Delta J}$ are constant. In addition, 
we also pay attention to the sub-peak whose position and relative height 
are shown as $t'_{\rm peak}$ and $r'$ in the insets in FIG. \ref{peakteff}(d) 
and FIG. \ref{rteff}(b), respectively.
\par
First we pay attention to the data at $\Delta T=0.26$K under $\Delta T$ perturbation.
As shown in FIG. \ref{peakteff}(a), the values of $t_{\rm peak}$ almost merge into the reference curve 
except for the data at the longest $t_{\rm eff}+t_{\rm w}$, 
where the main peak for the longest $t_{\rm eff}$ should appear at time much longer than  
the present observation time window ($\sim 10^4$s).
The sub-peak cannot be observed, and $r$ scarcely decreases from 1 (FIG .\ref{rteff}). Therefore, all the data at $\Delta T=0.26$K under $\Delta T$ perturbation satisfy the criteria for the $W$ regime in Table1. Thus, the accumulative aging proceeds in this condition.
\par
Under the $\Delta J$ perturbation in addition to $\Delta T=0.26$K, the $t_{\rm peak}$ except for the longest $t_{\rm eff}$ satisfies Eq. (\ref{weakpeak}), and
the $t_{\rm peak}$, at the longest $t_{\rm eff}$, is shorter than $t_{\rm w}+t_{\rm eff}$ (FIG. \ref{peakteff}(a)).
The value of $r$ clearly decreases from 1 except for the shortest $t_{\rm eff}$ (FIG. \ref{rteff}).
This indicates that $\Delta J$ perturbation decreases the overlap length according to Eq. (\ref{over}).
Under $\Delta T+\Delta J$ perturbation with $\Delta T$=0.26K, thus, 
the system is {\it not} completely in the $W$ regime except for the shortest $t_{\rm eff}$, 
and the chaotic nature partially emerges.
Thus, this regime is in the $WC$ regime except for the shortest $t_{\rm eff}$,
at which the system belongs to the $W$ regime.
\par
Under both the perturbations with $\Delta T=0.40$K (see Fig \ref{peakteff}(b)),
the values of $t_{\rm peak}$ almost merge into the reference curve 
except for the data at the longest $t_{\rm eff}+t_{\rm w}$.
At the longest $t_{\rm eff}+t_{\rm w}$, the values of $t_{\rm peak}$ under both the perturbations 
merge together, but lie below the reference curve.
Under $\Delta T$ perturbation, $r$ is 
clearly lower than 1 except for the small $t_{\rm eff}$,
while $r$ under $\Delta T+\Delta J$ perturbation is clearly lower than 1 over all range of $t_{\rm eff}$ (FIG. \ref{rteff}).
The sub-peak cannot be observed under either perturbations.
Therefore, at $\Delta T=0.40$K under the $\Delta T$ perturbation, 
the system belongs to the $WC$ regime except for short $t_{\rm eff}$, 
at which the system belongs to the $W$ regime.
Under the $\Delta T+\Delta J$ perturbation, the system belongs to the $WC$ regime.
\par
Next, we turn to the data under the strongest perturbation ($\Delta T=1.05$K)
prior to the discussion of complex behavior under the medium perturbation ($\Delta T=0.69$K).
At $\Delta T=1.05$K, $t_{\rm peak}$ cannot be determined at $\Delta T=1.05$K because the $S$ curves are so flattened except for the short $t_{\rm eff}$ (see FIG. \ref{peakteff}(d)).
The value of $r$ rapidly decreases as $t_{\rm eff}$ increases
and becomes almost constant in the long $t_{\rm eff}$ region, in which
$r$ under both the perturbations merge (FIG. \ref{rteff}). 
The sub-peak under $\Delta T$ perturbation is observed except for the shortest $t_{\rm eff}$ 
and becomes pronounced with increasing $t_{\rm eff}$ (see the inset of FIG.\ref{rteff}).
In the strongly perturbed regime,
the chaotic effect significantly emerges
and the main peak satisfies Eq. (\ref{strongpeak}). 
In addition, the sub-peak, which is attributed to the rejuvenation, is observed around the time necessary to stabilize the applied field, i.e., $t_{\rm s}\sim 100$s.
Under the perturbation with $\Delta T=1.05$K, thus, 
the system at long $t_{\rm eff}$ may be in the $S$ regime,
where the order parameter $\rho $ becomes saturated to a level common to both kinds of perturbations
and the rejuvenation effect becomes noticeable.
Except for the long $t_{\rm eff}$, 
the system would not necessarily be classified into the $S$ regime, 
because $\rho $ is not saturated although the sub-peak is observable.
Therefore, this system may belong to the $SC$ regime.
The system at the shortest $t_{\rm eff}$ under $\Delta T$ perturbation is 
classified into the $WC$ regime because of the absence of the sub-peak. 
\par
At $\Delta T=0.69$K, $t_{\rm peak}$ almost merges into the reference curve 
except for the data at the longest $t_{\rm eff}+t_{\rm w}$. At the longest $t_{\rm eff}$, however,
the $t_{\rm peak}$ under $\Delta T+\Delta J$ perturbation is shorter than that under $\Delta T$ perturbation.
Moreover, the sub-peak appears under $\Delta T+\Delta J$ perturbation only at the longest $t_{\rm eff}$ (see the inset of FIG. \ref{peakteff}).
Under both the perturbations with $\Delta T=0.69$K, $r$ rapidly decreases from 1 as $t_{\rm eff}$ increases, but does not become saturated (FIG. \ref{rteff}).
Under the $\Delta T+\Delta J$ perturbation with $\Delta T=0.69$K, thus, 
the system at the longest $t_{\rm eff}$  gets into the $SC$ regime, 
whereas it does not under the $\Delta T$ perturbation.
Under the $\Delta T$ perturbation with $\Delta T=0.69$K, 
the system belongs to the $WC$ regime except for the shortest observation time, 
at which it belongs to the $W$ regime because $r$ scarcely deviates from 1.
\par
To clearly evaluate the overlap length under both the perturbations, $r$ is abridged under the condition that $t_{\rm eff}$ is constant and $\Delta T$ is variable (FIG. \ref{rDT}). This mirrors the situation that  the domain size grown during the perturbation stage is fixed while the overlap length $L_{\Delta T+(\Delta J)}$ is varied.
At the shortest $t_{\rm eff}$ ($\sim 170$s) under $\Delta T$ perturbation, 
$r$ scarcely deviates from 1 as $\Delta T$ increases.
This indicates that $\tau_{\rm rec}$ is so short
that the order parameter in ghost domains easily recovers.
At long $t_{\rm eff}$, $r$ rapidly decreases as $\Delta T$ increases. This shows the crossover 
from the weakly to the strongly perturbed regimes through the decrease in the overlap length. 
The values of $r$ are smaller under $\Delta T+\Delta J$ perturbation than under $\Delta T$ perturbation,
but the difference in $r$ between the perturbations becomes indistinct as $\Delta T$ increases.
Ultimately it disappears at large $\Delta T$ due to the extremely long recovery time.
\par

\begin{table}[h]
\begin{center}
\caption{The defenition of classification of our data under various perturbation conditions.
(1) The condition $t_{\rm peak} = t_{\rm eff} + t_w$ is satisfied.
(2) Sub peak is observed.
(3) The condition $r \sim 1$ is satisfied.
(4) The value of r attains the saturation value.
For example, the data that belongs to the $W$ regime satisfy the conditon (1) and (3), does not satisfy (2) and (4).
}
\begin{tabular}{c|c c c c c|}
\hline\hline
Regime & (1) & (2) & (3) & (4) \\
\hline
 $W$ & Y & N & Y & N\\

 $WC$ & Y & N & N & N\\

 $SC$ & Y & Y & N & N\\

 $S$ & N & Y & N & Y\\
\hline\hline
\end{tabular}
\label{criteria}
\end{center}
\end{table}

\begin{figure}[t]
\includegraphics[width=6cm,keepaspectratio]{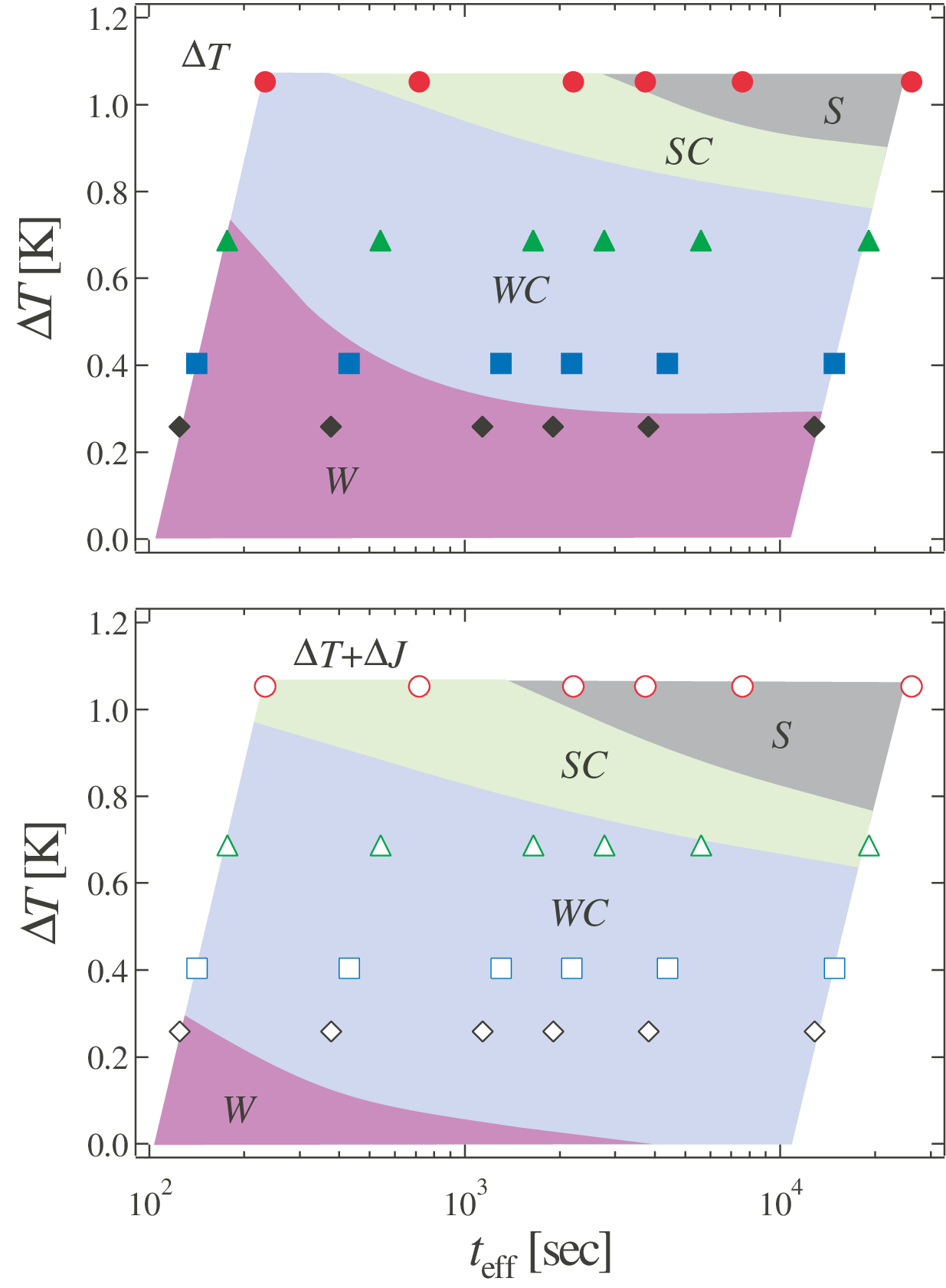}
\caption{(Color online) Schematic phase diagram where our data are classified into the four regimes:
the weakly perturbed regime ($W$ regime); 
the crossover regime, which is close to the weakly perturbed regime ($WC$ regime);
the crossover regime, which is close to the strongly perturbed regime ($SC$ regime);
the strongly perturbed regime ($S$ regime) (bottom to top).
The upper and lower figures show the diagram under $\Delta T$ and $\Delta T+\Delta J$ perturbations, respectively. 
}
\label{class}
\end{figure}

FIG. \ref{class} shows a schematic phase diagram 
in which the perturbation conditions are classified into the four regimes.
In this figure, the boundary curves are guides to help the eyes grasp the qualitative aspect.
As $t_{\rm eff}$ and $\Delta T$ increase, 
a systematic change from the $W$ to the $S$ regime is observed in the case of 
both the $\Delta T$ and  $\Delta T+\Delta J$ perturbations. 
In FIG. \ref{class}, we find a feature in which the boundary curve in the 
$\Delta T+\Delta J$ perturbation data lies below the corresponding boundary curve in the $\Delta T$ perturbation data. 
This can be interpreted in terms of the decrease in the overlap length due to the additional perturbation $\Delta J$.

\subsection{The effect of the bond perturbation}

The order parameter under $\Delta T+\Delta J$ perturbation are smaller than 
those under $\Delta T$ perturbation with the same $\Delta T$ as mentioned above. These indicate that $\Delta J$ perturbation decreases the order parameter
through the decrease in the overlap length.
This is clearly demonstrated in FIG. \ref{class} through the shift of boundary curves. 
The behavior of the sub-peak, observed under both perturbations with $\Delta T=1.05$K, 
suggests that the rejuvenation effect becomes pronounced due to the additional $\Delta J$, 
because the $t'_{\rm peak}$ is longer and $r'$ is larger under the $\Delta T+\Delta J$ perturbation 
than under the $\Delta T$ perturbation (see the inset of Fig. \ref{peakteff}(d) and Fig. \ref{rteff}). 
\par
In addition, we pay attention to the feature in which 
$r$ under $\Delta T+\Delta J$ perturbation with $\Delta T=0.26$K
is significantly smaller than $1$ while $r$ under $\Delta T$ perturbation 
is practically equal to $1$ as $t_{\rm eff}$ approaches zero, as shown in FIG. \ref{rteff}.
This suggests that the $\Delta J$ perturbation decreases the order parameter 
whereas the $\Delta T$ perturbation does not, even at infinitesimal $t_{\rm eff}$.
Thus, at $\Delta T=0.26$K, the effect of $\Delta T$ perturbation on the order parameter is so small
that the $\Delta J$ perturbation makes the dominant contribution. 
Based on the feature in which $r$ under $\Delta T+\Delta J$ perturbation with $\Delta T=$0.26K 
is practically equal to that under $\Delta T$ perturbation at $\Delta T= 0.40$K, 
we estimate that $\Delta J= 0.14\sim 0.40$K at $\Delta T=0.26$K. 

\section{Conclusion}
The bond perturbation $\Delta J$, under photo illumination, affects the aging behavior of
semiconductor spin glass.
We then estimated that $\Delta J\sim 0.14\sim 0.40$K at $\Delta T=0.26$K.  
Thus, the bond perturbation $\Delta J$ can significantly change the bond configuration, 
although the photo-induced magnetization in Cd$_{0.63}$Mn$_{0.37}$Te is negligible small.\cite{4kawai}
This effect {\it cannot} be explained in terms of the strong 
separation of the time scales on different length scales. 
We attribute it to the decrease in the overlap length, i.e. $L_{\Delta T+\Delta J}<L_{\Delta T}$.
Furthermore, we observed the crossover from weakly to strongly perturbed regimes 
in the bond cycling accompanied by the temperature cycling.
These experimental results strongly suggest that "chaos" and the overlap length, 
which are the key concepts in the droplet picture,
exist because the contribution of bond perturbation appears only in the overlap length.
\par
In the future, it will be necessary to conduct the "pure" bond cycling experiment under photo illumination
where there is no change in temperature. 
In addition, the mechanism of the bond perturbation using photo illumination
should be clarified.

\section*{Acknowledgment} 
This work was performed during the FY 2002 21st Century COE Program.
We would like to thank S. Yabuuchi and Y. Oba for their fruitful discussions with us.


\begin{thebibliography}{84}
\bibitem{CdMnTe_spinglass}
As shown in Fig.\ref{FCZFC}, the FC susceptibility increases as temperature decreases,
which is different from the conventional spin glass behavior. 
This is attributed to the paramagnetic contribution from the small magnetic clusters,
which are not belonging to the freezing spins. See Ref.22.
\end{thebibliography}
\end{document}